\documentstyle[aps,version2,preprint,epsfig]{revtex}

\begin{document}
\draft
%\preprint{OCIP/C 98-??}
%\preprint{hep-ph/9806xxx}
%\preprint{June 1998}
\begin{title}
A Quark Model Calculation of $\gamma\gamma\to 
\pi\pi$ Including Final State Interactions
\end{title}
\author{H.G. Blundell$^1$, S. Godfrey$^1$, G. Hay$^1$, and E.S. Swanson$^2$}
\begin{instit}
$^1$Ottawa-Carleton Institute for Physics \\
Department of Physics, Carleton University, Ottawa Canada K1S 5B6 \\
$^2$ Department of Physics and Astronomy, University of Pittsburgh, \\
Pittsburgh, PA 15260 and \\
Jefferson Laboratory, 12000 Jefferson Ave, Newport News, VA 23606
\end{instit}

\begin{abstract}
A quark model calculation of the processes $\gamma\gamma 
\to \pi^+\pi^-$ and $\gamma\gamma \to \pi^0\pi^0$ is performed.
At tree level, only 
charged pions couple to the initial state photons and neutral 
pions are not expected in the final state.  However, a small but 
significant $\gamma\gamma \to \pi^0\pi^0$ cross section is observed.  
We demonstrate that this may be accounted for by a rotation in isospin
space induced by final state interactions.
The resulting $\pi^+\pi^-$ cross section is in good 
agreement with experiment while the $\pi^0\pi^0$ cross section is in 
qualitative 
agreement with the data.  
\end{abstract}
%\pacs{PACS numbers: }

\section{Introduction}

The constituent quark model has proven exceptionally 
successful in describing the 
properties of hadrons \cite{qm}.
 Nevertheless it is universally 
acknowledged that the quark model spectroscopic analysis is overly 
simplistic, in part because  
couplings of resonances to other hadronic states are not included.  
In particular, the 
quark model at its most basic level does not include final state 
interactions \cite{fsi,messiah} 
and the effects of coupling resonances to decay channels 
\cite{weinstein96}.  One can view the constituent quark model as 
representing the lowest components of a Fock space expansion while 
final state interactions and coupled channel effects represent higher 
order contributions.  It is possible for these contributions to 
shift significantly the resonance pole position.

In this paper we study the contributions of final state interactions 
to the process $\gamma\gamma \to \pi \pi$ for relatively low center of 
mass energies.  In this energy regime the photons have sufficiently 
long wavelength that they do not resolve the pions into their 
constituent quarks.  This assertion has been formalized in the Low 
theorem \cite{felow},
which states that scalar QED is adequate to describe 
$\gamma \gamma \to \pi^+\pi^-$ near threshold.
However, scalar QED cannot explain the observed 
finite cross section to neutral pions.  We will find that including 
final state interactions gives qualitative agreement with the data.  
This is nontrivial in view of the subtlety of the reaction, the expected
importance of relativistic effects, and the simplifying assumptions we
have made.
Essentially, the $\pi^0\pi^0$ final state arises from the difference 
in the I=0 and I=2 $\pi\pi$ potentials which contribute to the 
rescattering of the final state pions in $\gamma\gamma \to \pi\pi$.  
Although the final state interaction contributions have been studied 
previously in the quark model\cite{bdi}, 
that initial study used the Fermi approximation to final state interactions,
did not take into account the differences in the I=0 and I=2 
potentials (which are crucial to describing the $\pi^0\pi^0$ final 
state), and employed rescaled pion potentials. As a result,
the model of Ref. \cite{bdi} predicts a substantial enhancement in charged
pion production and fails when applied to neutral pion production\cite{penn}.
Our conclusions differ because we have avoided these pitfalls.

Before proceeding we note that this process has been studied in other 
approaches which  give good agreement with experiment.  In 
particular, a recent calculation using chiral perturbation theory to two 
loops \cite{cpt} agrees with the data as do calculations using
dispersion relations with phase shift data from $\pi\pi$ scattering 
and constraints from unitarity, analyticity, and crossing \cite{disp}.
Our purpose here is therefore not so much to explain 
the data but to expand the regime of applicability of the quark model 
and to gain some experience so 
that we may apply these methods to situations where the 
techniques just mentioned are inapplicable.   Indeed, a large body
of data exists on vector-vector production from $\gamma\gamma$. Much
of this data is poorly understood and has led to wide spread 
speculation \cite{spec}. We believe that the results presented here
will aid in clarifying this enigmatic situation.

We begin in section II by summarizing the scalar QED results for 
$\gamma\gamma \to \pi^+ \pi^-$.  In 
particular, we expand the cross section in terms of partial wave 
amplitudes.  The expressions for the cross sections 
including final state interactions are derived in section III.  
In section IV we present 
our results and, finally, a summary is given in section V.

\section{The Scalar QED cross section for $\gamma\gamma\to \pi^+\pi^-$}

At low energies, long wavelength photons do not probe the 
quark constituents of pions. Thus the pions may be rigorously treated as 
point particles interacting via scalar QED \cite{felow}.
The cross section for the process $\gamma\gamma\to 
\pi^+\pi^-$ may be derived from the amplitudes shown in Fig. 1 and is 
given by

\begin{equation}
\sigma = \frac{\pi\alpha^2}{4m^2}\left[2x(1+x)\sqrt{1-x} - x^2(2-x)
\ln\left(\frac{1+\sqrt{1-x}}{1-\sqrt{1-x}}\right)\right], \label{ggpptotcross}
\end{equation}
where $x\equiv 4m^2/s$, $s=E_{cm}^2$ is the Mandelstam 
variable, and $m$ is the mass of the pion.  
Partial waves amplitudes, which are required when considering
final state interactions, are obtained from the helicity amplitudes:
\begin{eqnarray}
{\cal M}_{++}={\cal M}_{--} &=& \frac{8\pi \alpha \,x}{1-(1-x)\cos^2{\theta}},
\nonumber \\
{\cal M}_{\pm \mp} &=& \frac{8\pi \alpha \:(1-x) \sin^2{\theta} \;
e^{\pm 2i\phi}}{1-(1-x)\cos^2{\theta}},
\end{eqnarray}
where  $\theta$ and $\phi$ are the spherical polar coordinates of the 
outgoing $\pi^+$  and 
the subscripts $+$ and $-$ are the helicities of the two incoming 
photons.  Partial wave amplitudes may be extracted as follows:
\begin{equation}
f_{LM_L}^{\lambda_1 \lambda_2} = \int \!{\rm d}\Omega \;Y_{LM_L}^*
(\theta,\phi)\:
{\cal M}_{\lambda_1 \lambda_2}. \label{partwaves}
\end{equation}
One finds that $f_{LM_L}^{++}=f_{LM_L}^{--}$ is only non-zero for $M_L=0$ and
that $f_{LM_L}^{+-}=f_{L\,-\!M_L}^{-+}$ is only non-zero for
$M_L=2$.  The cross section is then written in terms of the partial waves as
\begin{equation}
\sigma = \frac{{\cal S}\sqrt{1-x}}{128\pi^2 s}
\left[ (f_{00}^{++})^2 + \sum_{L \geq2,
{\rm even}}^\infty \left[(f_{L0}^{++})^2 + (f_{L2}^{+-})^2 \right] \right]
\label{ggppcross}
\end{equation}
where ${\cal S}
\equiv 1/(1+\delta_{\pi\pi})$ is a statistical factor required for
the $\pi^0 \pi^0$ channel (to be discussed below).
The necessary  partial waves are given by:
\begin{eqnarray}
f_{00}^{++}&=& 8\pi^\frac{3}{2}\alpha \frac{x}{\sqrt{1-x}}
\ln\left(\frac{1+\sqrt{1-x}}{1-\sqrt{1-x}}\right), \nonumber \\
f_{20}^{++}&=& 4\sqrt{5} \pi^\frac{3}{2}\alpha\frac{x}{1-x} \left[
-6+\frac{2+x}{\sqrt{1-x}}\ln\left(\frac{1+\sqrt{1-x}}{1-\sqrt{1-x}}\right)
\right], \nonumber \\
f_{22}^{+-}&=& 4\sqrt{\frac{15}{2}} \pi^\frac{3}{2}\alpha
\left[\frac{10}{3}-\frac{2}{1-x} +\frac{x^2}{(1-x)^\frac{3}{2}}
\ln\left(\frac{1+\sqrt{1-x}}{1-\sqrt{1-x}}\right)
\right], \nonumber \\
f_{40}^{++}&=& 3\pi^\frac{3}{2}\alpha \frac{x}{1-x}\left[\frac{110}{3}-
\frac{70}{1-x}+\frac{3x^2+24x+8}{(1-x)^\frac{3}{2}}
\ln\left(\frac{1+\sqrt{1-x}}{1-\sqrt{1-x}}\right)\right], \nonumber \\
f_{42}^{+-}&=& 3\sqrt{10}\pi^\frac{3}{2}\alpha
\left[\frac{-54}{5}+\frac{76}{3(1-x)}-\frac{14}{(1-x)^2}+\frac{x^2(6+x)}
{(1-x)^\frac{5}{2}}
\ln\left(\frac{1+\sqrt{1-x}}{1-\sqrt{1-x}}\right)
\right], \nonumber \\
f_{60}^{++}&=& 8\sqrt{13}\pi^\frac{3}{2}\alpha \frac{x}{1-x}
\left[-\frac{231}{40}+
\frac{7(4-15x)}{8(1-x)}
-\frac{21(1+5x+5x^2)}{8(1-x)^2} \right.\nonumber \\
&&\left.+\frac{16+120x+90x^2+5x^3}
{16(1-x)^\frac{5}{2}}\ln\left(\frac{1+\sqrt{1-x}}{1-\sqrt{1-x}}\right)
\right].
\end{eqnarray}

\section{Final State Interactions}

Final state interactions (FSI) are additional interactions between particles 
in the final state that are not included in the basic interaction.  In 
our problem the basic interaction is the scalar QED process described 
in the previous section and the FSI's are purely hadronic and are
described by effective 
potentials between the outgoing particles.  We use the distorted wave 
Born approximation in our calculations \cite{messiah}, 
which relies on nonrelativistic 
quantum mechanics.  We include relativistic phase space where 
possible and use the relativistic scalar QED result for the basic 
interaction.  This mixture of nonrelativistic and relativistic 
elements is typical of the quark model.

The first step in including FSI's is to find the wavefunction 
describing the relative motion of the final-state particles due to the 
FSI potentials alone.  We choose to do this by solving a two
body Schr\"odinger equation with an effective $\pi\pi$ interaction.
%It is found using the 
%Schr\"odinger equation for a central force potential 
%
%\begin{equation}
%\left[\nabla^2_{\vec{r}}-2\mu V(r)+k^2\right]\psi(\vec{r})=0 \label{swe}
%\end{equation}
%
%where $\mu$ is the reduced mass of the particles, $V(r)$ is the potential 
%between the two particles and $k$ is the relative momentum between the 
%particles.  In the absence of a 
%potential , $k^2\equiv 2\mu E$, where $E$ is the kinetic energy of the 
%system.  
%After choosing the solution for a particular orbital angular 
%momentum and rewriting the wavefunction in spherical coordinates 
%$\psi_{lm}(\vec{r}) \equiv u_l(r) \,Y_{lm}(\Omega_r)$
The scattered waves are written in terms of spherical waves as
\begin{equation}
\psi_{\vec{k}}^-(\vec{r}) = 4 \pi \sum_{l=0}^\infty
\sum_{m_l=-l}^l i^l e^{-i\delta_l} \,u_l(k,r) \,Y_{lm_l}^*(\Omega_k) 
\,Y_{lm_l}(\Omega_r) 
\end{equation}
where $u_l(k,r)$, the radial solution of the Schr\"{o}dinger equation,
has the asymptotic form
\begin{equation}
u_l(k,r) \stackrel{r\rightarrow \infty}{\longrightarrow} \frac{1}{kr}
\sin{(kr-{\scriptstyle\frac{1}{2}}l\pi+\delta_l)}.
\end{equation}
Here $k$ is the relative momentum between the particles, 
$k^2=2\mu E$, where $\mu$ is the reduced mass 
of the particles and $E$ is the total energy of the system.

%The $\pi\pi$ potentials used to derive the final state wavefunction 
%are taken from Refs. \cite{swanson92,li94}. These potentials were
%using techniques developed by Barnes and Swanson\cite{swanson91}.  
%obtained by equating the T-matrix for the leading order 
%interactions in perturbation theory for the nonrelativistic quark 
%model to the T-matrix for point-like mesons.  The quark model potentials 
%include one-gluon-exchange and the confining potential followed by 
%rearrangement into color singlet for I=0 and I=2. 
%For the I=0 
%channel there is an additional contribution from $q\bar{q}$ 
%annihilation.  The effective potentials are parametrized in the form 

The $\pi\pi$ potentials used to derive the final state wavefunction are
taken from Refs. \cite{li94,swanson91}. These are obtained by calculating
the perturbative T-matrix for a given process in the nonrelativistic
quark model. This T-matrix was then equated to a model pointlike
T-matrix to obtain an effective interaction. The quark model calculation
includes one gluon exchange and confinement interactions followed by
quark rearrangement into final state color singlets. 
For the I=0 
channel there is an additional contribution from $q\bar{q}$ 
annihilation.  The resulting effective potentials are parametrized in the form 

\begin{equation}
V_{\pi\pi}(r)=V_0\:e^{ -\frac{1}{2}\left(\frac{r}{r_0}\right)^2}.
\label{potparam1}
\end{equation}
The parameters, $V_0$ and $r_0$, are reproduced in Table 1.
%and the potentials are shown in Fig. 2.  
FSI corrections were only made to the S-wave amplitudes because 
corrections to D-waves 
were found to have a  negligible effect on the cross section.
%We only include results for FSI corrections to the $L=0$ partial wave of
%$\gamma \gamma \to \pi \pi$ because the $L=2$ potentials 
%are smaller than those for $L=0$, and we found that their effect
%on the cross section was negligible.

%To take into account FSI's we replace the partial wave amplitudes in 
%eqn. 4 with amplitudes corrected for FSI's.  
%To make the FSI corrections to the QED amplitudes we use 
%nonrelativistic collision theory in the Born approximation 
%\cite{messiah} which gives

Final state interactions are incorporated with the standard two-potential
relation \cite{fsi,messiah}:
\begin{equation}
\langle f|T|i\rangle ^{\rm FSI} 
= \langle \psi_{\vec{k}_f}^- | W | \phi_{\vec{k}_i} \rangle 
= \frac{1}{(2\pi)^{\frac{3}{2}}}\int\!{\rm d}^3\vec{k} \;\psi_{\vec{k}_f}^{-*}(\vec{k})\:
\langle f|T|i\rangle
\end{equation}
where $\psi_{\vec{k}_f}^{-*}(\vec{k})$ is the complex conjugate of the
momentum-space solution of the Schr\"{o}\-dinger equation with
the FSI potential $V_{\pi\pi}$, $W$ is the basic electromagnetic interaction,
and $\phi_{\vec k}$ is a plane wave.  The corrected QED amplitude 
is 
\begin{eqnarray}
{\cal M}^{\rm FSI}(\vec{k}_f)&=& \frac{2}{\pi} \sqrt{s(k_f)} \:
\sum_{l=0}^{\infty} \sum_{m_l=-l}^l 
e^{i\delta_l} \:Y_{lm_l}(\Omega_{k_f}) 
 \nonumber \\
&&\times 
\int\!{\rm d}^3\vec{k} \:\frac{{\cal M}(\vec{k})}{\sqrt{s(k)}} 
\:Y_{lm_l}^*(\Omega_k)
\int_0^\infty \!{\rm d}r \:r^2 j_l(kr) \:u_l(k_f,r) \nonumber
\end{eqnarray}
where ${\cal M}(\vec{k})$ is the uncorrected scalar QED amplitude, 
$j_l(kr)$ is a spherical Bessel function and
the  $\sqrt{s}$ factors are introduced to relate the 
relativistic QED amplitudes to the nonrelativistic amplitudes used in 
the calculation.  

In order to include the FSI correction for each partial wave we decompose 
the amplitudes
\begin{equation}
{\cal M}^{FSI}(\vec{k})\equiv \sum_{L=0}^{\infty}
 \,\sum_{M_L=-L}^L f_{LM_L}^{FSI}(s(k)), 
Y_{LM_L}(\Omega_k)
\end{equation}
where the corrected partial wave amplitudes are given by
\begin{equation}
f_{LM_L}^{\rm FSI}(s(k_f))=\frac{2}{\pi} \sqrt{s(k_f)} \;e^{i\delta_L} 
\int_0^\infty \!{\rm d}k \int_0^\infty \!{\rm d}r \:r^2\: k^2 \:
\frac{f_{LM_L}(s(k))}{\sqrt{s(k)}} \,j_L(kr) \:u_L(k_f,r).
\end{equation}
Helicity labels have been omitted.
In the absence of an FSI potential, $\delta_L\rightarrow 0$ and
$u_L(k_f,r)\rightarrow j_L(k_f\,r)$, the orthogonality of the 
spherical
Bessel functions yields a delta function, and we recover 
$f_{LM_L}^{\rm FSI}(s(k_f)) \rightarrow f_{LM_L}(s(k_f))$ as expected.

The relationship between the charged pion basis of the physical scalar QED
reaction and the isospin basis of the potentials is
\begin{eqnarray}
|\pi^+\pi^-\rangle &=& \sqrt{\frac{2}{3}}\:|00\rangle+
\sqrt{\frac{1}{3}}\:|20 \rangle, \nonumber \\
|\pi^0\pi^0\rangle &=& -\sqrt{\frac{1}{3}}\:|00\rangle+
\sqrt{\frac{2}{3}}\:|20 \rangle. \nonumber
\end{eqnarray}
%Thus, starting with the expression for the FSI corrected amplitude
%\begin{equation}
%\langle f|T|i\rangle ^{\rm FSI} = 
%\frac{1}{(2\pi)^3}\int\!{\rm d}^3\vec{k} \;\langle \psi_{\vec{k}_f}
%^{-\pi\pi}|\phi_{\vec{k}}^{\pi^+\pi^-} \rangle \:\langle \phi_{\vec{k}}
%^{\pi^+\pi^-} |W|\phi_{\vec{k}_i} \rangle 
%\end{equation}
%the first bra-ket is transformed to the isospin basis as
Thus, the physical pion wavefunctions can be written in terms of the 
momentum space solutions of the 
Schr\"odinger equation in the isospin basis:
\begin{eqnarray}
\psi^{-\pi^+\pi^- *}(k) =
\langle \psi_{\vec{k}_f}^{-\pi^+\pi^-}|\phi_{\vec{k}}^{\pi^+\pi^-} \rangle
 &=& 
\left[ \sqrt{\frac{2}{3}}\:\langle \psi_{\vec{k}_f}^{-0}|+\sqrt{\frac{1}{3}}
\:\langle \psi_{\vec{k}_f}^{-2}| \right]
\left[ \sqrt{\frac{2}{3}}\:|\phi_{\vec{k}}^0\rangle+\sqrt{\frac{1}{3}}\:
|\phi_{\vec{k}}^2 \rangle \right]
\nonumber \\
&=&(2\pi)^{\frac{3}{2}} \left[\frac{2}{3}\: \psi_{\vec{k}_f}^{-0*}(\vec{k}) +
\frac{1}{3}\: \psi_{\vec{k}_f}^{-2*}(\vec{k}) \right], \nonumber \\
\psi^{-\pi^0\pi^0 *} =
\langle \psi_{\vec{k}_f}^{-\pi^0\pi^0}|\phi_{\vec{k}}^{\pi^+\pi^-} \rangle
 &=&
(2\pi)^{\frac{3}{2}} \left[-\frac{\sqrt{2}}{3}\:\psi_{\vec{k}_f}^{-0*}(\vec{k})
+\frac{\sqrt{2}}{3}\: \psi_{\vec{k}_f}^{-2*}(\vec{k}) \right]. \nonumber
\end{eqnarray}
%depending on which pion final state is produced.  
The corrected partial waves for $\pi^+\pi^-$ production are thus given by
\begin{eqnarray}
f_{LM_L\,\pi^+\pi^-}^{\rm FSI}(s(k_f))&=&\frac{2}{\pi} \sqrt{s(k_f)}  
\int_0^\infty \!{\rm d}k \int_0^\infty \!{\rm d}r \:\:r^2\: k^2 \:
\frac{f_{LM_L}(s(k))}{\sqrt{s(k)}}  \nonumber \\
&& \times\,j_L(kr) \left[\frac{2}{3}\: 
e^{i\delta_L^0} \:u_L^0(k_f,r)+\frac{1}{3}\: e^{i\delta_L^2}\:
u_L^2(k_f,r) \right] 
\label{eq:fsiamp}
\end{eqnarray}
with an analogous expression for the $\pi^0\pi^0$ final state. Here
$\delta_L^I$ is the $\pi\pi$ phase shift in the isospin $I$, partial wave $L$
channel.
To simplify the following expressions we define the isospin amplitudes
\begin{equation}
g_{LM_L}^I(s(k_f))\equiv \frac{2}{\pi} \sqrt{s(k_f)}  
\int_0^\infty \!{\rm d}k \int_0^\infty \!{\rm d}r \;r^2\: k^2 \:
\frac{f_{LM_L}(s(k))}{\sqrt{s(k)}} \,j_L(kr)\:u_L^I(k_f,r).
\end{equation}
This yields the final state enhanced probability for $\pi^+\pi^-$ production:
\begin{equation}
\left|f_{LM_L\,\pi^+\pi^-}^{\rm FSI}\right|^2 = \frac{4}{9} \left(g_{LM_L}^0
\right)^2+\frac{1}{9}
\left(g_{LM_L}^2\right)^2+\frac{4}{9} \:g_{LM_L}^0\: g_{LM_L}^2 \,
\cos{(\delta_L^0-\delta_L^2)} 
\end{equation}
and for $\pi^0 \pi^0$ production:
\begin{equation}
\left|f_{LM_L\,\pi^0\pi^0}^{\rm FSI}\right|^2 = \frac{2}{9} \left(g_{LM_L}^0
\right)^2+\frac{2}{9}
\left(g_{LM_L}^2\right)^2-\frac{4}{9} \:g_{LM_L}^0\: g_{LM_L}^2 \,
\cos{(\delta_L^0-\delta_L^2)} .
\label{eqn:fisp0p0}
\end{equation}
With no $\pi\pi$ potentials $g_{LM_L} \to f_{LM_L}$ 
and $\delta^I_L=0$ and the original cross sections are recovered.

The last step is to account for the limited polar acceptance of the $\pi\pi$
experimental data.  Integrating over a finite solid angle 
($-\cos{\theta_{\rm acc}} \leq \cos{\theta} \leq \cos{\theta_{\rm
acc}}$) to obtain the uncorrected cross section for limited polar 
acceptance is straightforward:  Eq.~\ref{ggpptotcross} is replaced by
\begin{eqnarray}
\sigma_{\rm acc} &=& \frac{\pi\alpha^2}{4m^2}\left[2\cos{\theta_{\rm acc}}\;
x\sqrt{1-x}\left(\frac{x^2}{1-(1-x)\cos^2{\theta_{\rm acc}}}+1 \right) 
\right. \nonumber \\
&& \left. - x^2(2-x) \ln\left(\frac{1+\sqrt{1-x}\,\cos{\theta_{\rm acc}}}
{1-\sqrt{1-x}\,\cos{\theta_{\rm acc}}}\right)\right].
\end{eqnarray}

However, the expression in terms of partial waves is affected more 
drastically as the spherical harmonics are no longer orthogonal to 
each other and 
the contributions of the partial waves to the total cross section 
can no longer be separated.
The expression replacing 
Eq.~\ref{ggppcross} becomes an infinite sum.
%(where we have dropped the helicity labels)
%
\begin{eqnarray}
\sigma_{\rm acc} &=& \frac{\sqrt{1-x}}{256\pi^2 s}
\int_0^{2\pi} \!\!{\rm d}\phi \int_{-\cos{\theta_{\rm acc}}}
^{\cos{\theta_{\rm acc}}} \!\!{\rm d}(\cos{\theta}) \,
\left[ 2\left| \sum_{L \geq 0,{\rm even}}^\infty f_{L0} \,Y_{L0}(\theta,\phi)
\right|^2  \right. \nonumber \\
&& + \left. \left| \sum_{L \geq 2,{\rm even}}^\infty f_{L2}
\,Y_{L2}(\theta,\phi)
\right|^2+ \left| \sum_{L \geq 2,{\rm even}}^\infty f_{L2} \,
Y_{L\,-\!2}(\theta,\phi)\right|^2 \right]. \label{fsiinftyterms}
\end{eqnarray}

Because the $L=0$ partial wave is the only one affected
significantly by FSI's, we assume that all of the observed $\gamma \gamma \to
\pi^0 \pi^0$ events are in an $L=0$ state.  Since that distribution is 
spherically symmetric, we can correct the data by
simply dividing by the value of $\cos{\theta_{\rm acc}}$ appropriate to the
experiment in question.

For $\gamma \gamma \to \pi^+ \pi^-$ the total observed cross section is a
mixture of all partial waves. We again assume that only the $L=0$ wave is 
affected by FSI's.  To simplify our calculations we define the 
harmonic integrals
\begin{equation}
h_{l_1l_2}^m \equiv \int_0^{2\pi} \!\!{\rm d}\phi 
\int_{-\cos{\theta_{\rm acc}}}
^{\cos{\theta_{\rm acc}}} \!\!{\rm d}(\cos{\theta}) \:Y_{l_1m}^*(\theta,\phi)
\,Y_{l_2m}(\theta,\phi)
\end{equation}
and note that $h_{l_1l_2}^m = h_{l_2l_1}^m$ and $h_{l_1l_2}^{-\!2}=
h_{l_1l_2}^{2}$ ($h$ is independent of $m$)..  We can then write the total 
cross section with only the $L=0$
partial wave corrected for FSI's as the total uncorrected cross section, plus
an infinite number of correction terms, each of which is the difference between
the corrected and uncorrected values of a term in Eq.~\ref{fsiinftyterms}.
Only those terms involving $f_{00}$, which goes to $f_{00\,\pi^+\pi^-}^{\rm
FSI}$, need correcting.  Eq.~4 is replaced by
\begin{eqnarray}
\lefteqn{\sigma^{\rm FSI}_{{\rm acc},\pi^+\pi^-} = 
\sigma_{\rm acc} +
\frac{\sqrt{1-x}}{128\pi^2 s}\left[ \rule[-.5cm]{0cm}{1.0cm}
h_{00}^0\left\{\left|f_{00\,\pi^+\pi^-}^{\rm FSI}\right|^2 -(f_{00})^2 \right\}
\right.} \nonumber \\
&&\left. +\sum_{L \geq 2,{\rm even}}^\infty h_{L0}^0 \left\{
f_{00\,\pi^+\pi^-}^{\rm FSI} \,f_{L0} + 
\left(f_{00\,\pi^+\pi^-}^{\rm FSI}\right)^* f_{L0} 
-2 f_{00} f_{L0} \right\}\right] \nonumber \\
&=&  \sigma_{\rm acc} +
\frac{\sqrt{1-x}}{128\pi^2 s}\left[ \rule[-.5cm]{0cm}{1.0cm}
h_{00}^0\left\{\frac{4}{9} \left(g_{00}^0
\right)^2+\frac{1}{9}
\left(g_{00}^2\right)^2+\frac{4}{9} \:g_{00}^0\: g_{00}^2 \,
\cos{(\delta_0^0-\delta_0^2)}-(f_{00})^2 \right\}
\right. \nonumber \\
&&\left. +\sum_{L \geq 2,{\rm even}}^\infty 2\,h_{L0}^0 \:f_{L0} \left\{
\frac{2}{3} \,g_{00}^0\:\cos{\delta_0^0} +
\frac{1}{3} \,g_{00}^2\:\cos{\delta_0^2} 
- f_{00} \right\}\right]. 
\end{eqnarray}
Excellent convergence over the relevant energy range is obtained
when the series is truncated after $L=6$.  Fortunately, the three
experiments whose data we use for comparisons all have the same limited polar
acceptance ($|\cos{\theta}| \leq 0.6$), so the  results need be corrected 
for only one value of $\cos{\theta_{\rm acc}}$.

\section{Results}

Equation (13) demonstrates that the isospin dependence of the $\pi\pi$
production amplitudes is generated through the isospin dependence
of the final state interactions. The fact that the isoscalar and 
isotensor $\pi\pi$ interactions differ markedly causes a rotation in
isospin space which is reflected in a nonzero $\pi^0\pi^0$ amplitude
(see Equation (15)). Indeed, strong, but isospin independent, 
$\pi\pi$ interactions would {\it not} give rise to neutral pions in the
final state. 

Equation (15) reveals that the details of the neutral pion prediction
depend on the scalar QED production mechanism, the isoscalar and isotensor 
hadronic phase shifts, and FSI-induced modifications to the outgoing pion-pion
wavefunction. We shall assume that the tree order scalar QED amplitude
captures most of the production mechanism (we note that direct $\gamma\gamma$
coupling to the $f_0$ and rescattering to $\pi^0\pi^0$ is possible -- this is
discussed below) and that
the joining of relativistic and nonrelativistic formalisms has not
overly distorted the predictions. 

The S-wave elastic phase shifts produced by the $\pi\pi$ potentials are 
shown as solid
lines in Figures 3 ($I=2$) and 4 ($I=0$). It is clear that 
both underestimate the strength of the $\pi\pi$ interaction.  However, it
has been shown \cite{swanson92} that the majority of this discrepancy may
be alleviated by employing relativistic kinematics in the description of
the $\pi\pi$ phase shift. This yields the dashed curves in Figures 3 and 4. 
The agreement in the isotensor sector is quite good. The isoscalar prediction
is still somewhat low; however, the isoscalar sector differs from the isotensor
sector in that $q\bar q$ annihilation will lead to resonance coupling. 
Indeed, the authors of Ref. \cite{li94} have shown that
very good agreement may be achieved once the neglected effects of mixing with 
the $f_0(980)$ and $f_0(1350)$ are included. Since resonance coupling will
also induce wavefunction effects in our problem, we have chosen to
examine the importance of the phase shifts by simply fitting them. The fits
are indicated as dotted lines in Figures 3 and 4. The differences between
the predictions for neutral pion production for the relativized model 
(dashed lines) and the fits (dotted lines) will be examined below.

The final contribution to the $\pi^0\pi^0$ cross section came from the 
FSI-induced distortions to the pion-pion wavefunction. We shall assume that
this is accurately described by our potential formalism. This may appear
doubtful because the potential phase shifts did not agree well with elastic
$\pi\pi$ scattering data. However, as noted above, the majority
of the discrepancy is due to the nonrelativistic phase space necessitated
by the numerical solution of the Schr\"odinger equation, rather than the
potential itself. We are thus confident that the effective potentials
can describe the wavefunction distortion. However, the model neglects 
additional distortions to the wavefunction caused by transitions to virtual
resonance states (the $f_0$). For the moment, we will assume that these are
small near threshold since the $f_0(980)$ is narrow and the $f_0(1350)$ is
distant. Indeed, calculations with a relativistic Breit-Wigner yield
the following cross sections for $\pi^0\pi^0$ at threshold: 0.015 nb for
$f_0(980)$, 0.00005 nb for $f_0(1275)$, and 0.15 nb for $f_2(1275)$.

Figure 5b shows the predicted cross section for $\gamma\gamma \to \pi^+\pi^-$
along with data from three experiments. There is very little difference 
between predictions, indicating the lack of importance of FSI effects in the
charged channel.  The Born result agrees fairly well with the data. 
The Born result also agrees well with the dispersion analysis of 
Ref. \cite{disp} for $\sqrt{s} < 350$ MeV, where there is no data. 

The dashed line of Figure 5a is the predicted cross section for $\pi^0\pi^0$
production which is based on the relativized $\pi\pi$ phase shift predictions
of Ref. \cite{swanson91} and a nonrelativistic $\pi\pi$ final state 
wavefunction. Recall that this curve therefore does not incorporate 
resonance effects and does not fit the
isoscalar $\pi\pi$ scattering data well. The result presented in Figure 5a
agrees well with data near threshold -- an indication that the near
threshold neutral pion data can be explained by the postulated mechanism
of final state interaction-induced isospin rotation. Nevertheless, the dashed
curve falls below the data by a factor of three or four above energies of
350 MeV. It is therefore of interest to test the effects of intermediate
resonance states. We model this by fitting the experimental S-wave $\pi\pi$
phase shifts as described above. Note that we are ignoring wavefunction
distortion caused by dynamical resonance effects. The resulting prediction is
given by the dotted line in Figure 5a. The agreement with data is 
substantially improved. This is due to the increased accuracy of the
description of the $I=0$ scattering data and is a strong indication that
resonance effects (as manifested in the $I=0$ phase shift) can be important even near threshold. 

The agreement
with experiment is not perfect -- we believe that this is likely due to 
explicit dynamical resonance effects in the final state scattering 
wavefunction. It should be possible to test this assertion by including 
resonances in a coupled channel description of the final state interactions.
Nevertheless, the main point is established -- it is possible to describe
the rather complicated processes leading to a $\pi^0\pi^0$ final state
with a sufficiently detailed analysis of the production mechanism
and the final state interactions.

\section{Conclusions}

The reaction $\gamma\gamma \to \pi^0\pi^0$ presents an interesting
theoretical challenge. Neutral pions do not couple directly to low
energy photons so that direct techniques based on the Low theorem
cannot be used to describe the cross section. Calculations tend
to be difficult; this is 
illustrated by efforts in chiral perturbation theory which 
have found it necessary to introduce three new counterterms, work to two
loops, and to unitarize. We have presented a relatively simple explanation
of the data. In our model neutral pion production proceeds via charged
pion production followed by final state hadronic interactions. The 
$\pi^+\pi^-$ initial state is described well by the Born order
scalar QED amplitude, in accord with Low's theorem. This amplitude is
then convoluted with the final state pion-pion wavefunction to arrive
at a prediction for $\pi^0\pi^0$ production.

The result for $\gamma\gamma \to \pi^+\pi^-$ is very insensitive to 
final state interactions -- a result that follows from isospin algebra
and the weakness of the final state interactions \cite{ess}. However,
the prediction for $\pi^0\pi^0$ depends crucially on the interference
between $I=0$ and $I=2$ final state interactions, and is therefore
driven by the well known differences in these processes -- $I=2$ S-wave 
$\pi\pi$ scattering is repulsive while $I=0$ is attractive. It was
shown in Ref. \cite{li94} that roughly half of the scattering strength
at low energy in $I=0$ is due to the $f_0(980)$ and $f_0(1350)$ resonances
-- even at threshold.  
This leads to the surprising conclusion that resonance coupling is important
to getting the magnitude of the $\pi^0\pi^0$ cross section correct near 
threshold.

This calculation was undertaken with an eye to extensions to heavier
systems. In particular, $\gamma\gamma$ production of vector-vector states
has a long, and enigmatic, history. For example, the large ratio
$\sigma(\gamma\gamma \to \rho^0\rho^0)/\sigma(\gamma\gamma \to \rho^+\rho^-)$
is problematic and has engendered much speculation. This is due, is part,
to the lack of clear theoretical guidance --  
chiral perturbation theory and dispersion methods are inapplicable for these
processes.
Of course such constraints do not exist on the quark model and we look
forward to applying it to an explication of this interesting area of
hadronic physics.

\acknowledgments

The authors are grateful to Ted Barnes for extensive discussions on 
$\gamma \gamma$ physics.
This research was supported in part by the Natural Sciences and Engineering 
Research Council of Canada and by the United States Department of 
Energy under grant DE-FG02-96ER40944 and contract DE-AC05-84ER40150 
under which the Southeastern Universities Association operates the
Thomas Jefferson National Accelerator Facility.

\figure{The tree-level Feynman diagrams for the scalar QED 
contributions to $\gamma\gamma \to \pi^+\pi^-$.  The labels denote the 
particle momenta $(p_i)$ and photon polarizations $(\epsilon_i)$.}

\figure{The $\pi\pi$ potentials vs $r$ for $I=0$, $L=0$ (solid 
line), $I=0$, $L=2$ (dashed line), $I=2$, $L=0$ (dotted line), and
$I=2$, $L=2$ (dot-dot-dashed line).  The potentials are given by 
Eqn.~\ref{potparam1} with the parameters given in Table I.}

\figure{The
$I=2$ $\pi\pi$ scattering phase shift vs.\ $\protect\sqrt{s}$
using $\pi\pi$ potentials (solid line),
the quark model prediction \cite{li94} (dashed line), 
and a linear approximation to the experimental data, 
$\delta^{I=2}_{L=0}=-0.00062 \; \hbox{rad/MeV} \times 
(\sqrt{s}-250 \; \hbox{MeV})$ (dotted line).  
The data are from Hoogland {\it et al.} (circles) \
\cite{hoogland} -- we show the results of both of their methods; 
and Prukop {\it et al.} (triangles)
\ \cite{prukop} -- we show the results of 
their first fit.
Some of the data shown in this
figure were obtained from References~\cite{durhamRALHEP} and
\cite{morgan94}.  The horizontal error bars on the data indicate
bin size; for the vertical error bars all of the given errors were added
in quadrature.}

\figure{The
$I=0$ $\pi\pi$ scattering phase shift vs.\ $\protect\sqrt{s}$, 
using $\pi\pi$ potentials (solid line),
the quark model prediction without resonance contributions (see text)
\cite{li94} (dashed line), 
and a linear approximation to the experimental data, 
$\delta^{I=0}_{L=0}=0.0027 \; \hbox{rad/MeV} \times 
(\sqrt{s}-250 \; \hbox{MeV})$ (dotted line).  
The data are from Mukhin {\it et al.} (circles) \
\cite{mukhin}; Rosselet {\it et al.} (inverted triangles)
\ \cite{rosselet} --
the horizontal bars only approximate their bins, and their data is actually 
for $\delta_0^0-\delta_1^1$ -- we have used the $\delta_0^0$ data
extracted from it by Li {\it et al.}\ \cite{li94}; Estabrooks and 
Martin \cite{estabrooks} -- we show the results of both their s- 
(squares) and 
t-channel fits (diamonds); 
and Protopopescu {\it et al.} (triangles) \ \cite{protopopescu} -- we 
show the results of their case 1.  
For further comments see Fig. 3.}

\figure{(a) $\sigma(\gamma
\gamma \to \pi^0 \pi^0)$ vs $\protect\sqrt{s}$, using phase shifts
{}from the quark model  (dashed line),
{}from  linear fits to the data (dotted line),
 and experimental data.  The data are
{}from Bienlein ({\it et al.}\ ) (closed circles)
\cite{bienlein}, Marsiske {\it et al.}\ (open circles) \cite{marsiske} and
Edwards {\it et al.}\ (closed triangles) \cite{edwards}.  
The data have been corrected
to full polar acceptance, $|\cos{\theta}| \leq 1.0$.  
(b) $\sigma(\gamma \gamma \to \pi^+ \pi^-)$ vs.\ $\protect\sqrt{s}$
with the same line labelling as (a).  The curves have been corrected
to have a limited polar acceptance to match the data: $|\cos{\theta}| \leq
0.6$.   
The data are from Behrend {\it et al.}\ (closed circles)
\cite{behrend}, Boyer {\it et al.}\ (open circles) \cite{boyer}, and 
Aihara {\it et al.}\ (closed triangles) \cite{aihara}.  
For additional comments see Figure 3. }

\newpage

%\vskip 0.2cm
%\begin{table}[t]
%\vspace{-0.3cm}
\noindent
{\bf Table 1:} 
{\small The parameters of the $\pi-\pi$ potentials used in this
work \cite{swanson92,li94}.  The
t-channel gluon exchange potentials have two contributions, due to
colour-hyperfine and confinement terms, which must be summed.  The potentials
use the parametrization of Eqn.~\ref{potparam1} 
with $V_0$ in GeV, $r_0$ in ${\rm GeV}^{-1}$.}
\vskip 0.2cm
\begin{center}
\begin{tabular}{|c|c||c|c|c|c|c|c|} \hline
%\multicolumn{2}{|c||}{} & \multicolumn{6}{c|}{$\pi$--$\pi$ Potentials 
%($V_0$ in GeV, $r_0$ in ${\rm GeV}^{-1}$)} \\ \cline{3-8}
\multicolumn{2}{|c||}{Pion State} & 
\multicolumn{4}{c|}{t-channel gluon exchange} & 
\multicolumn{2}{c|}{s-channel} \\ \cline{1-6}
 & & \multicolumn{2}{c|}{hyperfine} & 
\multicolumn{2}{c|}{confinement} & 
\multicolumn{2}{c|}{ gluon exchange} \\ \cline{3-8}
\hspace{0.25cm}L\hspace{0.25cm} & I & $V_0$ & $r_0$ & $V_0$ & $r_0$ & $V_0$ & 
$r_0$ \\ \hline\hline
0 & 0 & $-0.392$ & 1.36 & $-0.024$ & 2.29 & $-1.75$ & 1.48 \\
0 & 2 & 0.786 & 1.36 & 0.047 & 2.29 &  & \\
2 & 0 & $-0.044$ & 1.40 & 0.0175 & 1.49 &  & \\
2 & 2 & 0.088 & 1.40 & $-0.035$ & 1.49 &  & \\\hline
\end{tabular}
\end{center}
\vspace{-0.4cm}
%\caption
%{The parameters of the potentials used in this
%work \cite{swanson92,li94,swanson96}.  The
%t-channel gluon exchange potentials have two contributions, due to
%colour-hyperfine and confinement terms, which must be summed.  The potentials
%are parametrized as in Eqs.~\ref{potparam1}.}
%\label{t:potparam}
%\end{table}
%
\vskip 0.4cm

\newpage

\vskip 0.4cm
\centerline{\epsfig{file=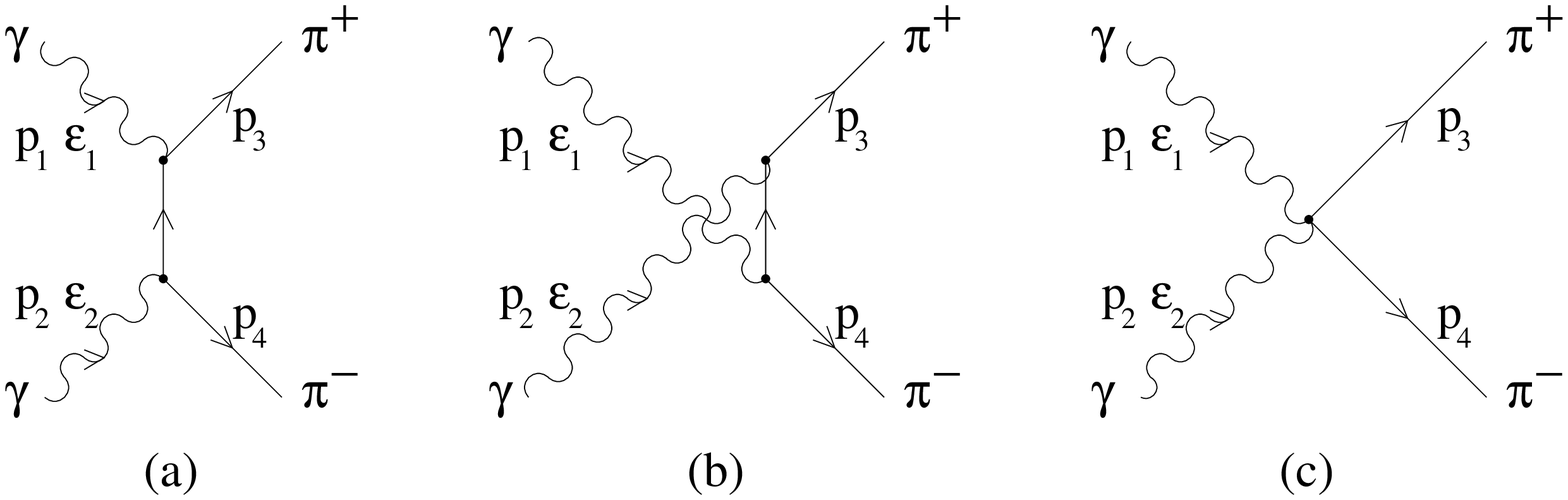,width=14.0cm,clip=}}
\noindent
{\bf Fig 1:} {\small The tree-level Feynman diagrams for the scalar QED 
contributions to $\gamma\gamma \to \pi^+\pi^-$.  The labels denote the 
particle momenta $(p_i)$ and photon polarizations $(\epsilon_i)$.}

\vskip 2.0cm

\vskip 0.2cm
\centerline{\epsfig{file=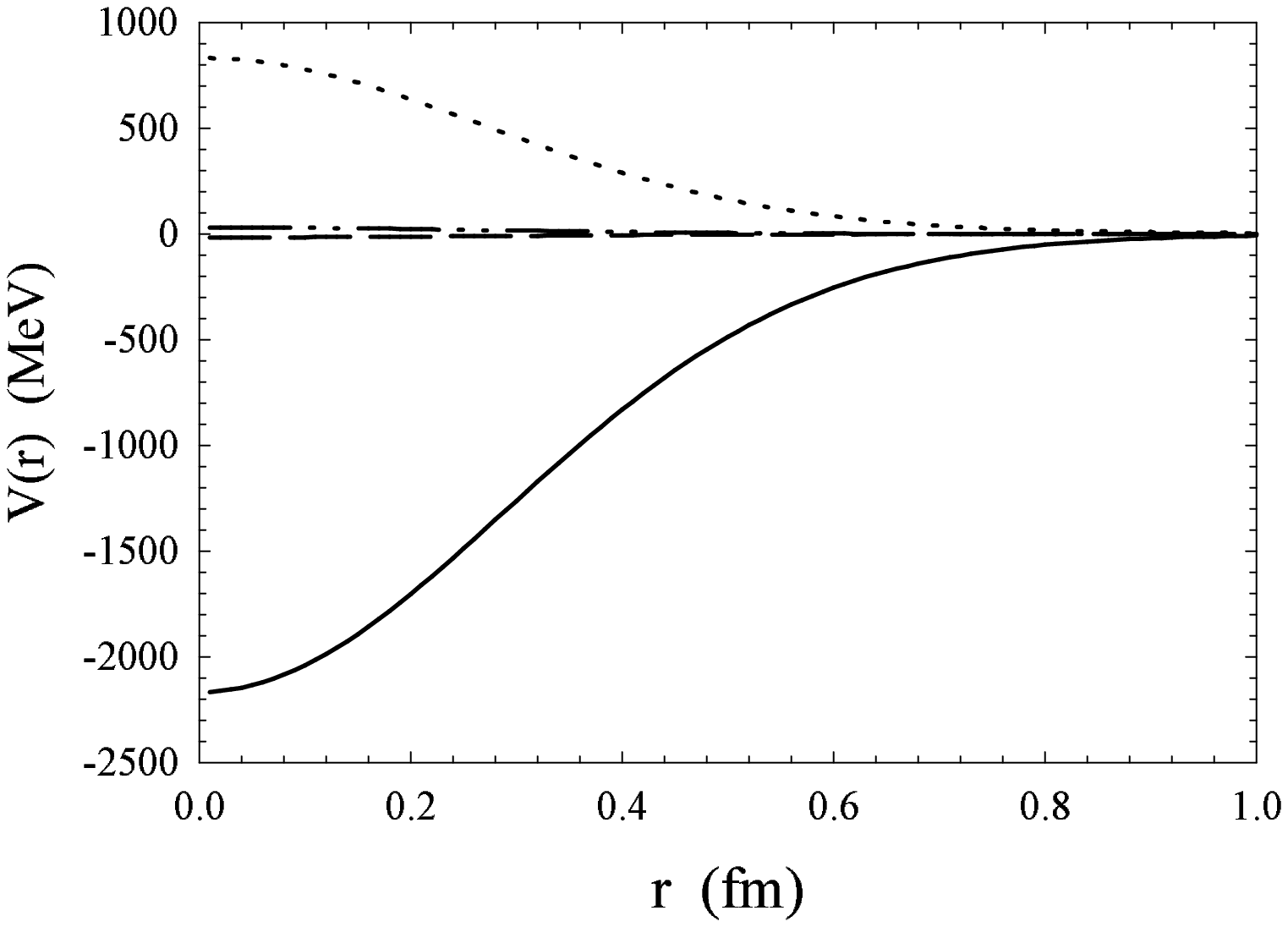,width=9.0cm,clip=}}

\noindent
{\bf Fig 2:} The $\pi\pi$ potentials vs $r$ for $I=0$, $L=0$ (solid 
line), $I=0$, $L=2$ (dashed line), $I=2$, $L=0$ (dotted line), and
$I=2$, $L=2$ (dot-dot-dashed line).  The potentials are given by 
Eqn.~\ref{potparam1} with the parameters given in Table I.

%\vskip 0.4cm

\newpage

%\begin{figure}[t]
%\vspace{-2cm}
%\begin{center}
%\makebox{\epsfxsize=6.5in\epsffile{ph-sh2.eps}}
%\end{center}
%\vspace{-1.3cm}
\vskip 0.4cm
\centerline{\epsfig{file=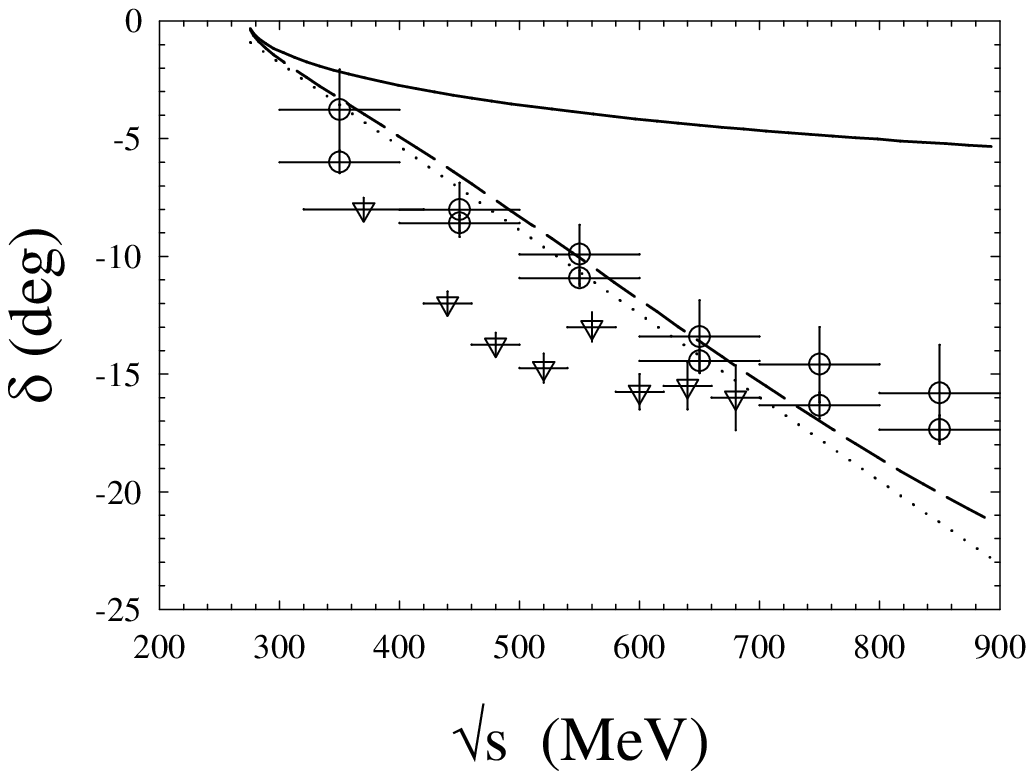,width=10.0cm,clip=}}

\noindent
{\bf Figure 3:}
%\caption
{\small The
$I=2$ $\pi\pi$ scattering phase shift vs.\ $\protect\sqrt{s}$
using $\pi\pi$ potentials (solid line),
the quark model prediction \cite{li94} (dashed line), 
and a linear approximation to the experimental data, 
$\delta^{I=2}_{L=0}=-0.00062 \; \hbox{rad/MeV} \times 
(\sqrt{s}-250 \; \hbox{MeV})$ (dotted line).  
The data are from Hoogland {\it et al.} (circles) \
\cite{hoogland} -- we show the results of both of their methods; 
and Prukop {\it et al.} (triangles)
\ \cite{prukop} -- we show the results of 
their first fit.
Some of the data shown in this
figure were obtained from References~\cite{durhamRALHEP} and
\cite{morgan94}.  The horizontal error bars on the data indicate
bin size; for the vertical error bars all of the given errors were added
in quadrature.
}
%\label{f:phases2}
%\end{figure}

\newpage

%\begin{figure}[t]
%\vspace{-2cm}
%\begin{center}
%\makebox{\epsfxsize=6.5in\epsffile{ph-sh-0.eps}}
%%\end{center}
%\vspace{-1.3cm}
\vskip 0.4cm
\centerline{\epsfig{file=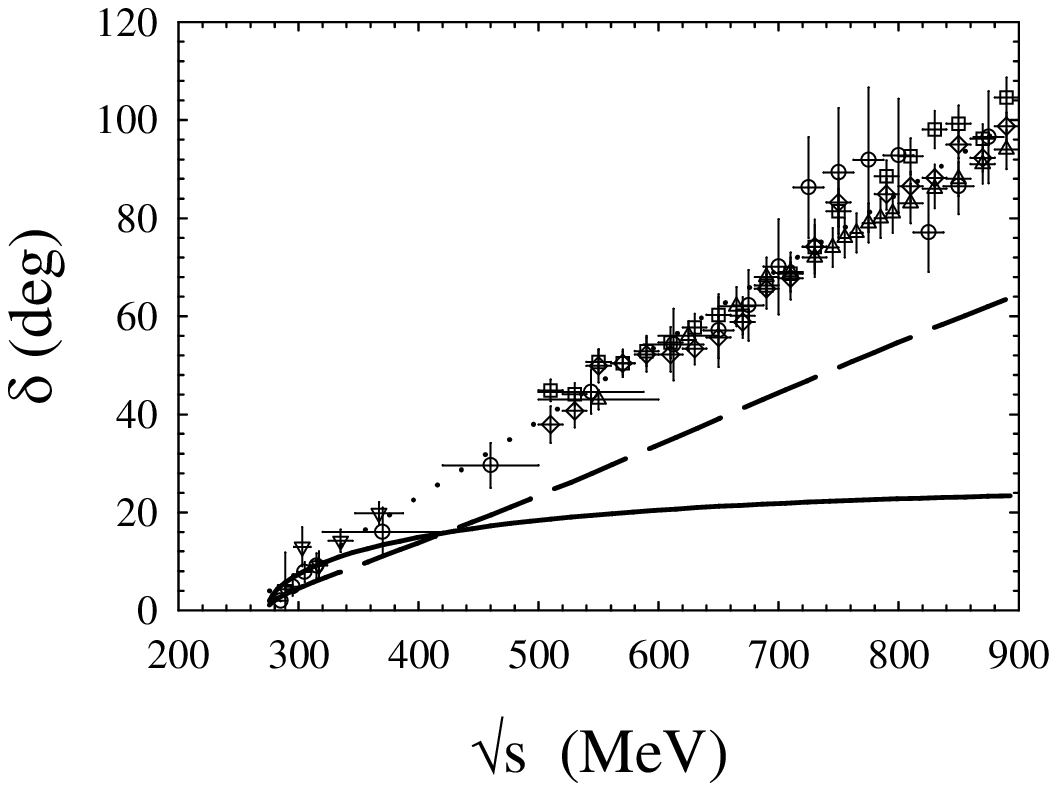,width=10.0cm,clip=}}

\noindent
{\bf Figure 4:}
%\caption
{\small The
$I=0$ $\pi$--$\pi$ scattering phase shift vs.\ $\protect\sqrt{s}$, 
using $\pi\pi$ potentials (solid line),
the quark model prediction without resonance contributions (see text)
\cite{li94} (dashed line), 
and a linear approximation to the experimental data, 
$\delta^{I=0}_{L=0}=0.0027 \; \hbox{rad/MeV} \times 
(\sqrt{s}-250 \; \hbox{MeV})$ (dotted line).  
The data are from Mukhin {\it et al.} (circles) \
\cite{mukhin}; Rosselet {\it et al.} (inverted triangles)
\ \cite{rosselet} --
the horizontal bars only approximate their bins, and their data is actually 
for $\delta_0^0-\delta_1^1$ -- we have used the $\delta_0^0$ data
extracted from it by Li {\it et al.}\ \cite{li94}; Estabrooks and 
Martin \cite{estabrooks} -- we show the results of both their s- 
(squares) and 
t-channel fits (diamonds); 
and Protopopescu {\it et al.} (triangles) \ \cite{protopopescu} -- we 
show the results of their case 1.  
For further comments see Fig. 3.
}
%\label{f:phases0}
%\end{figure}

\newpage

%\begin{figure}[t]
%\vspace{-2cm}
%\begin{center}
%\makebox{\epsfxsize=6.5in\epsffile{fig-4.eps}}
%\end{center}
%\vspace{-1.3cm}
\vskip 0.4cm
\centerline{\epsfig{file=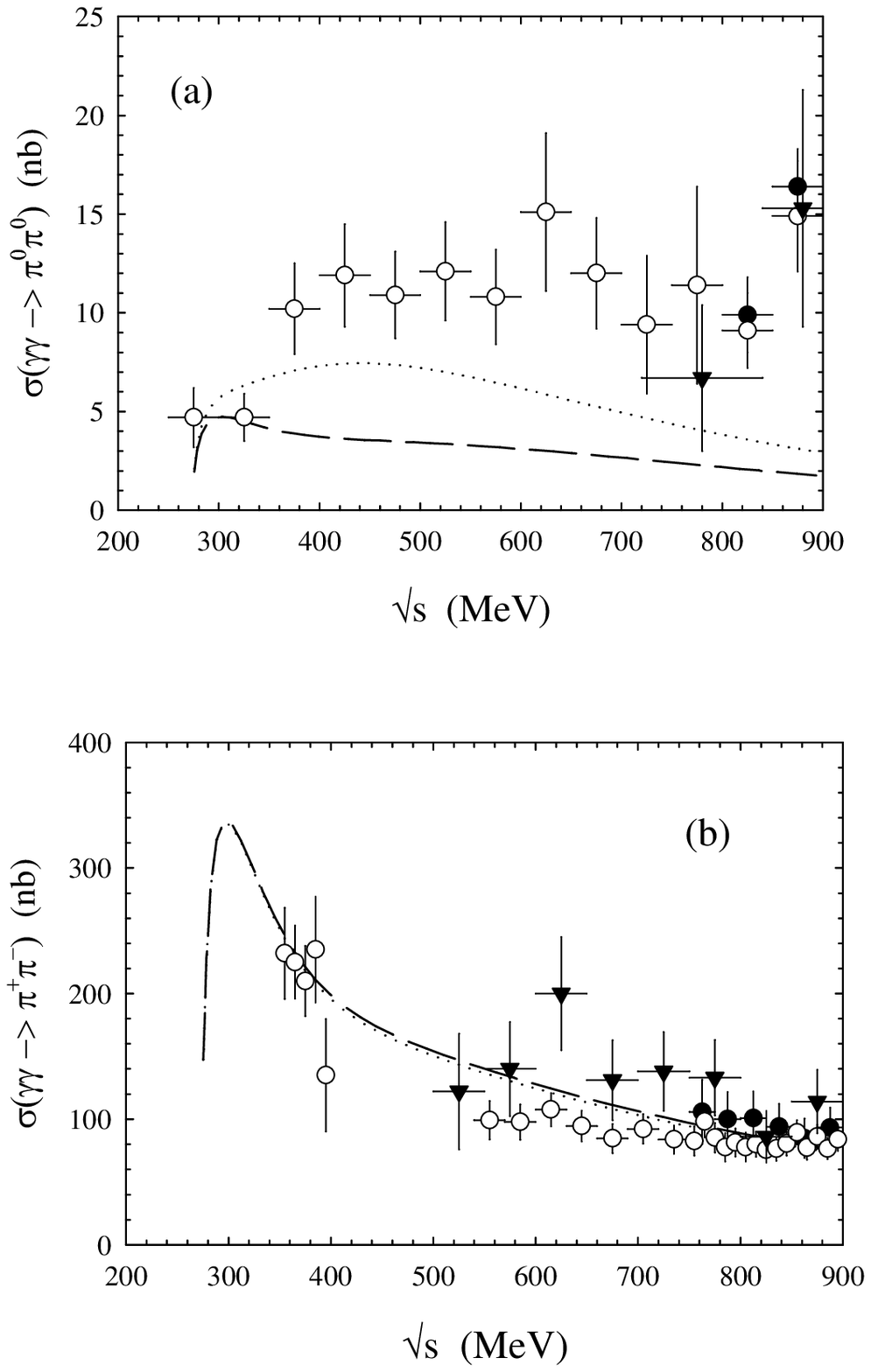,width=10.0cm,clip=}}

\noindent
{\bf Figure 5:}
%\caption
{\small (a) $\sigma(\gamma
\gamma \to \pi^0 \pi^0)$ vs $\protect\sqrt{s}$, using phase shifts
{}from the quark model  (dashed line),
 from  linear fits to the data (dotted line),
 and experimental data.  The data are
{}from Bienlein ({\it et al.}\ ) (closed circles)
\cite{bienlein}, Marsiske {\it et al.}\ (open circles) \cite{marsiske}, and
Edwards {\it et al.}\ (closed triangles) \cite{edwards}.  
The data have been corrected
to full polar acceptance, $|\cos{\theta}| \leq 1.0$.  
(b) $\sigma(\gamma \gamma \to \pi^+ \pi^-)$ vs.\ $\protect\sqrt{s}$
with the same line labelling as (a).  The curves have been corrected
to have a limited polar acceptance to match the data: $|\cos{\theta}| \leq
0.6$.   
The data are from Behrend {\it et al.}\ (closed circles)
\cite{behrend}, Boyer {\it et al.}\ (open circles) \cite{boyer}, and 
Aihara {\it et al.}\ (closed triangles) \cite{aihara}.  
For additional comments see Figure 3.
}
%\label{f:ggppcross001}
%\end{figure}

\end{document}